\documentclass[aps,floats,twocolumn,prd,nofootinbib,superscriptaddress]{revtex4-2}

\usepackage{fnpct} 
\usepackage{comment}
\usepackage{ifpdf}
\usepackage{slashed}
\usepackage{color}
\usepackage[mathscr]{eucal}
\usepackage[utf8]{inputenc}
\usepackage{cancel}
\usepackage{bm}
\usepackage{ulem}
\usepackage{hyperref}

\usepackage{braket}

\usepackage[most]{tcolorbox}

\usepackage{graphicx} 
\graphicspath{{}} 

\tcbset{colback=white, colframe=black, 
        highlight math style= {enhanced, 
            colframe=red,colback=red!10!white,boxsep=0pt}
        }

\definecolor{red}{rgb}{1,0,0}
\definecolor{darkred}{rgb}{0.6,0,0}
\definecolor{darkgreen}{rgb}{0.992447,0.623778,0.034597}
\definecolor{ppink}{rgb}{1,0.4,0.4}
\definecolor{bblue}{rgb}{0.284602,0.317763,0.963947}
\definecolor{purple}{rgb}{0.5 ,0, 0.7}

\definecolor{dgreen}{rgb}{0 ,0.5, 0.5}

\definecolor{cbPurple}{RGB}{204,121,167}    

\makeatletter
\newcommand\footnoteref[1]{\protected@xdef\@thefnmark{\ref{#1}}\@footnotemark}
\makeatother

\allowdisplaybreaks[1]

\begin{document}

\title{
Dynamics of self-gravitating halos with decaying dark matter
}

\author{Anna Bencke}
\email{abencke1@jhu.edu}
\affiliation{
 William H.\ Miller III Department of Physics \& Astronomy, Johns Hopkins University, 3400 N.\ Charles St., Baltimore, MD 21218, USA
}

\author{Marc Kamionkowski}
\affiliation{
 William H.\ Miller III Department of Physics \& Astronomy, Johns Hopkins University, 3400 N.\ Charles St., Baltimore, MD 21218, USA
}

\author{Annika H. G. Peter}
\affiliation{Department of Physics, The Ohio State University, 191 W.\ Woodruff Avenue, Columbus, Ohio 43210, USA
}
\affiliation{Department of Astronomy, The Ohio State University, 141 19th Ave. W., Columbus, Ohio 43210, USA
}
\affiliation{Center for Cosmology and AstroParticle Physics (CCAPP), The Ohio State University, 191 W. Woodruff Ave., Columbus, Ohio 43210, USA
}

\author{Juniper Morales}
\affiliation{
 Baltimore Polytechnic Institute, 1400 West Cold Spring Lane, Baltimore, MD 21209}

\begin{abstract}
We study the evolution of self-gravitating spherical dark matter halos composed of dark matter particles that decay to a slightly less massive particle and a massless particle. The small kick velocity imparted to the daughter particle upon decay heats the halo.  If the kick velocity is small compared with the characteristic velocity, the halo remains bound but puffs out. We also consider, for the first time, decaying halos with anisotropic initial conditions and find non-negligible effects on the resulting density profile. To simulate the halo evolution, we utilize the recently developed \texttt{NSphere} code across a broad range of decay times and kick velocities. We train an emulator to provide a mapping between the decay parameters and the relaxed density profile. We discuss how the resulting halo-scale signatures complement existing cosmological constraints on decaying dark matter and can be tested by upcoming surveys.
\end{abstract}

\maketitle

\section{Introduction}

The existence of dark matter is inferred from a variety of independent probes, including the flat rotation curves of spiral galaxies~\cite{Rubin1980}, the dynamics of galaxy clusters~\cite{Zwicky1933}, and strong and weak gravitational lensing measurements~\cite{Clowe2006}. Kinematic observations of visible matter indicate that dark matter is distributed in extended halos surrounding galaxies and galaxy clusters, with density profiles that have been extensively characterized through numerical simulations~\cite{NFW1996,NFW1997}. Despite this observational evidence, the fundamental nature of dark matter remains unknown. While the canonical cold dark matter paradigm is broadly consistent with observations on large scales, a variety of small-scale discrepancies have motivated the exploration of dark matter models with properties beyond minimal gravitational interactions~\cite{Bullock2017,Buckley:2017ijx,Tulin2018,Nadler:2026waz}.

One such extension involves dark matter particles that undergo decay to warm daughter products~\cite{Aoyama:2011, Aoyama2014}. In this framework, we postulate a cold dark matter particle $\chi$ with mass $M_\chi$ and lifetime $\tau$ that decays into a less massive daughter particle $\Upsilon$ with mass $M_\Upsilon = M_\chi(1 - \epsilon)$ and a massless, non-interacting particle $\zeta$. Upon decay, the daughter particle $\Upsilon$ receives a kick velocity $v_{\rm kick} \simeq \epsilon c$, where $\epsilon \ll 1$. The dynamical response is governed by the ratio of the kick velocity to the characteristic velocity $v_s$: when $v_{\rm kick} \gg v_s$, a substantial fraction of the decay products acquire enough velocity to escape the gravitational potential, leading to halo dissolution \cite{Peter_2010}. The observable consequences also depend on the lifetime $\tau$, with the halo structure largely preserved when $\tau$ exceeds the age of the universe. 

Two-body decaying dark matter (DDM) models have received attention in the context of tensions in the $\Lambda$CDM paradigm. Through free-streaming of the decay products, DDM suppresses the growth of small-scale structure, potentially reconciling measurements of the amplitude of matter fluctuations $S_8$ \cite{Peter_2010,Enqvist:2015, FrancoAbellan:2020xnr, FrancoAbellan:2021, Alvi:2022aam, Acharya:2026bmo, Nygaard:2023gel, Holm:2022eqq}. Cosmological probes place upper bounds on the kick velocity; CMB data bound $v_{\rm kick} \lesssim 300$--$3000\,{\rm km\,s^{-1}}$ for lifetimes of order the age of the universe \cite{Bucko:2022kss, Bucko:2024izb, Bencke:2026uws}, while the one-dimensional Lyman-$\alpha$ forest flux power spectrum imposes $v_{\rm kick} \lesssim 30$--$70\,{\rm km\,s^{-1}}$ for lifetimes $\lesssim 10\,{\rm Gyr}$ \cite{Wang:2013rha, Wang2014, Fuss:2022zyt, Zhao:2026wxi}. Nonlinear probes, including weak lensing, the halo mass-concentration relation, and Milky Way satellite counts, can be sensitive to kick velocities $\lesssim 100\,\rm{km\,s}^{-1}$ for lifetimes slightly longer than the age of the universe \cite{Peter2010b, Wang:2012eka, Peter:2010sz,DES:2022doi}. Thus, the nonlinear halo response provides a complementary, and potentially stronger, probe of DDM physics.

Previous investigations of halo evolution in the presence of DDM have been constrained by numerical limitations, as the full $N$-body simulations that have been used are computationally expensive. Recently, the \texttt{NSphere} code has been developed to efficiently evolve self-gravitating spherical systems by exploiting Newton's shell theorem, wherein the gravitational force on any particle depends only on the enclosed mass and is directed radially~\cite{Kamionkowski:2025uae}. Under the assumption of spherical symmetry, the six-dimensional phase-space distribution reduces to three effective dimensions, enabling systematic exploration of the parameter space governing DDM halos.

In this work, we consider a range of lifetimes and kick velocities within the small-scale free-streaming bounds of interest. We allow the halos to evolve into a new relaxed state, and discuss the effects on the density profile and velocity dispersions, which we find to be consistent with \cite{Nadler:2025yni}. Introducing initially anisotropic distribution functions, we find that the impact on the relaxed density profiles is small, but there is some interesting physics regarding the effect of anisotropy on the evolution of the density profile at small radius. We train an emulator on a grid of DDM parameters, and find that it can interpolate the final relaxed density profile to within a median fractional error of $\sim 1\%$. This emulator can also be used to derive the enclosed-mass profiles of halos with DDM, for which we provide an analytic fit.

This paper is organized as follows. In Section~\ref{sec:methods}, we outline the phenomenology of decaying dark matter particles and describe the implementation in \texttt{NSphere}. Section~\ref{sec:results} presents distribution profiles obtained across a range of decay lifetimes and kick velocities for isotropic initial conditions, and Section~\ref{sec:anisotropy} introduces anisotropic distribution functions. The emulator is detailed in Section~\ref{sec:emulator}. Finally, Section~\ref{sec:conclusions} summarizes our findings and discusses cosmological implications.

\section{Methods}
\label{sec:methods}

In this section, we introduce the \texttt{NSphere} code \cite{Kamionkowski:2025uae} and discuss how it is augmented to include decaying dark matter.

\subsection{The \texttt{NSphere} code}

The first key insight underlying \texttt{NSphere} is that with spherical symmetry, the gravitational force on any particle is directed radially and depends only on the mass at smaller radii. The second is that each orbit lies in a plane, but the distribution of those planes is isotropic, and so the six-dimensional phase space of a general collisionless system can be reduced to three effective dimensions: the radial distance $r$ from the halo center, the speed $v$, and the angle $\theta$ between the velocity vector and the radial direction. The halo is represented as a collection of $N$ concentric spherical shells, each characterized by its radius $r_n$, radial velocity $v_n$, and specific angular momentum $\ell_n = r_n v_n \sin\theta_n$. The equation of motion for each shell is
\[
\ddot{r} = -\frac{G M(r)}{r^2} + \frac{\ell^2}{r^3},
\]
where $G$ is Newton's constant, and $M(r)$ denotes the enclosed mass at radius $r$. At each timestep $\Delta t$, particles are sorted by radius, enabling the enclosed mass to be computed directly from the radial rank of each shell. The positions and velocities are then advanced using a leapfrog integrator.  An adaptive refinement scheme is employed for particles on low-angular-momentum orbits, for which the equations of motion become stiff near pericenter.
This algorithmic structure yields a computational complexity of $\mathcal{O}(N \ln N)$, dominated by the sorting operation. Furthermore, because the $N$ particles populate a three-dimensional rather than six-dimensional phase space, statistical fluctuations are suppressed relative to full $N$-body realizations with comparable particle number, and spurious two-body relaxation is eliminated.

We take for the initial conditions a spherical halo with a given density profile and a self-consistent velocity distribution.  For our primary analysis, the velocities are chosen from an isotropic velocity distribution, but we then also explore anisotropic velocity distributions.

\subsection{Implementation of dark matter decays}

To add a decaying component to the \texttt{NSphere} code, we give each particle a flag denoting whether or not it has decayed. Each decay is described by its decay lifetime $\tau$ (in Myr) and its velocity kick $v_{\rm kick}\,$ (in km s$^{-1}$). At each timestep, we compute the probability that a particle has decayed via $\Delta t /\tau$. We use a thread-safe random number generator to determine if the particle will decay in this timestep. If the particle decays, it is given a velocity kick with magnitude $v_{\rm kick}\,$ and a random, isotropic direction with respect to its rest frame.

In this paper, we consider Hernquist profiles, though we note that this analysis can easily be extended to other halo profiles (e.g. Navarro--Frenk--White or Einasto). The Hernquist profile is
\begin{eqnarray}
\label{eqn:hernquist}
    \rho(r) = \frac{\rho_s}{(r/r_s)(1+r/r_s)^3},
\end{eqnarray}
where $\rho_s$ is the scale density and $r_s$ is the scale radius, such that $c = R_{\rm vir}/r_s$ is the concentration of the halo given a virial radius $R_\text{vir}$. We define $R_{\rm vir}$ as the point where the mean density is 200 times the cosmological critical density. 

We present results everywhere in dimensionless Hernquist units, expressing density and radius in units of $\rho_s$ and $r_s$, respectively, and the kick velocity and decay time in units of the characteristic velocity and time scales
\[
v_s \equiv \left(\frac{GM}{r_s}\right)^{1/2},\qquad t_s \equiv \left(\frac{r_s^3}{GM}\right)^{1/2},
\]
where $M$ is the halo mass. These are the natural velocity and time units of the Hernquist model and should not be identified with the virial velocity or virial dynamical time of the halo. This dimensionless formulation allows our results to be rescaled to arbitrary halo mass and scale radius. For our fiducial choice of halo mass ($M = 10^{12}\,M_\odot$) and scale radius ($r_s=23$ kpc), representative of a Milky Way scale halo \cite{refId0}, the characteristic scales are $v_s = 432\,\rm{km\,s}^{-1}$ and $t_s = 52\,\rm{Myr}$. 

For decay lifetimes much longer than the simulation runtime and for velocity kicks much smaller than the virial velocity, the decay has a limited impact on the evolution of the halo. Conversely, a simulated halo with a short decay lifetime and a large velocity kick will experience rapid evaporation. Therefore, we hereafter limit our analysis to kick velocities in the range of $[0.01, 0.5]\,v_s$ and decay lifetimes in the range of $[0.5, 10]\,t_s$. The kick velocities we consider therefore span $v_{\rm kick} \approx 4$--$216\,{\rm km\,s^{-1}}$,  bracketing the small-scale free-streaming bounds from the Lyman-$\alpha$ forest \cite{Wang:2013rha, Wang2014, Fuss:2022zyt, Zhao:2026wxi}.

We use $10^6$ particles in all simulations. The decaying runs are evolved for 200 scale times with $10^5$ timesteps, at which the relaxed DDM profile is numerically converged. We allow for the simulations to run long enough for all decays to happen and for subsequent halo relaxation; for our choice of scale time, about 10 Gyr.

We also evolve a control halo with decays disabled, using the same initial conditions, particle number, and final time as the DDM simulations. The evolved control profile shows as a numerical artifact a reduction of the density in the inner halo, with the largest deviations occurring below approximately $0.5\,r_s$. These changes are common to both the control and DDM simulations. Consequently, comparisons intended to isolate the effects of decay are made relative to the evolved control profile $\rho_{\rm null}$, rather than directly relative to the initial Hernquist profile. Because the numerical relaxation of the central cusp is timestep dependent (Appendix~\ref{appendix:convergence}), the control is evolved at $10^6$ timesteps so that it provides a converged reference. We further verified that the ratio $\rho_{\rm DDM}/\rho_{\rm null}$ converges for a null simulation with $10^6$ timesteps and a DDM simulation with $10^5$ timesteps. This ratio isolates the physical decay response down to $r \simeq 0.1\,r_s$, below which residual finite-timestep bias sets the inner limit of the emulator's validity.

\section{Results}
\label{sec:results}

In this Section, we discuss our results, long-running DDM halo simulations using \texttt{NSphere} focusing on how the density profile and radial velocity dispersion evolve as functions of the decay parameters $v_{\rm kick}$ and $\tau$ and in time. The grids shown in Figs.~\ref{fig:density_profile_grid} and~\ref{fig:velocity_dispersion_grid} are an illustrative survey, chosen to display the qualitative transition from a negligible response at low kick velocity to strong structural evolution at high kick velocity; they span $0.01\le v_{\rm kick}/v_s\le0.4$ and $0.5\le\tau/t_s\le7.5$. The quantitative emulator of Sec.~\ref{sec:emulator} is instead trained on a denser $10\times10$ grid restricted to $0.1\le v_{\rm kick}/v_s\le0.5$, the range over which the density modification rises above the numerical baseline.

\subsection{Density profiles}

\begin{figure}
    \centering
    \includegraphics[width=\linewidth]{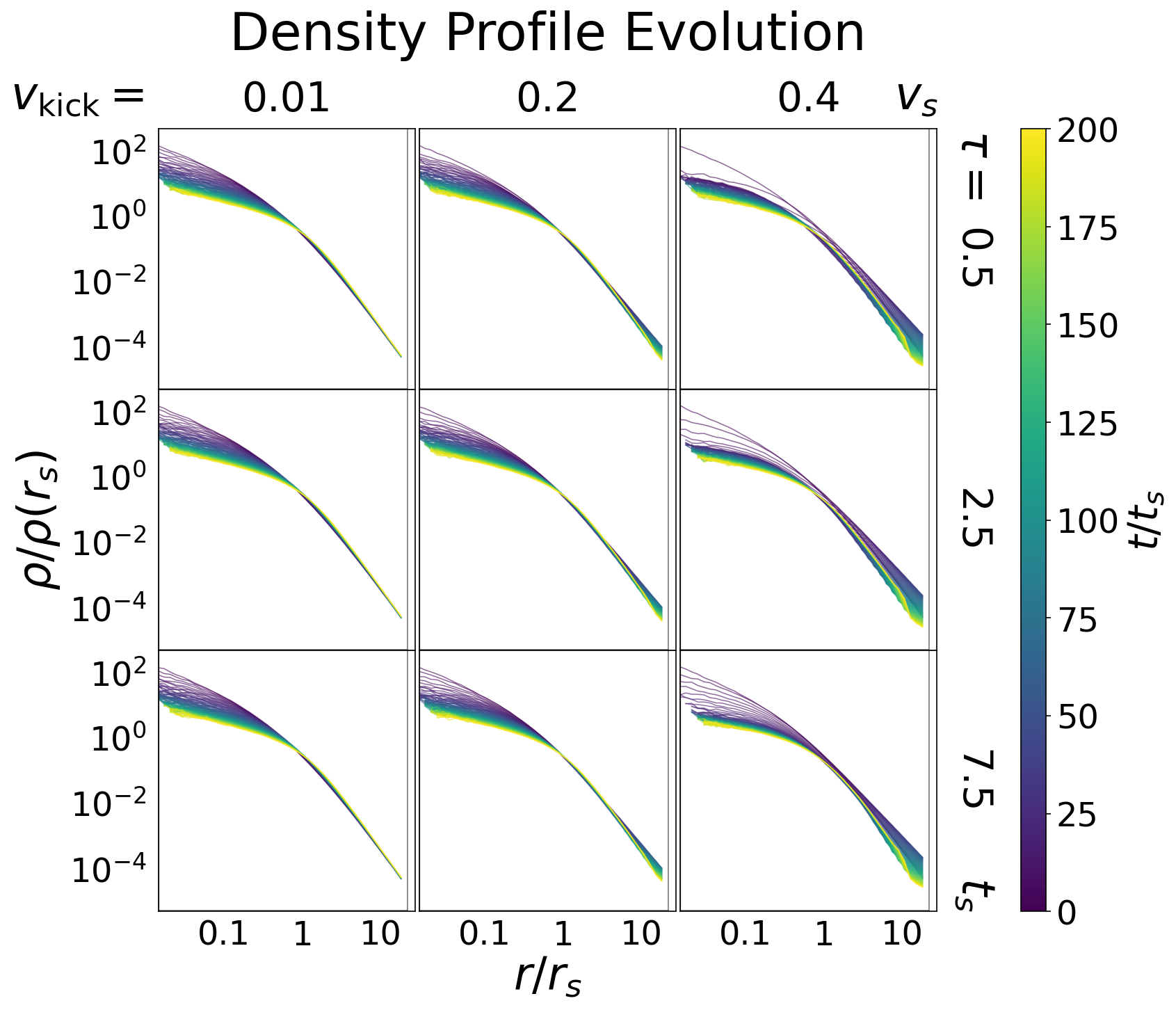}
    \caption{Density profile evolution of Hernquist profile halos with decaying dark matter. The kick velocity $v_{\rm kick}$ ranges from $0.01$ to $0.4$ in units of the scale velocity, increasing from left to right. The decay lifetime $\tau$ ranges from $0.5$ to $7.5$ in units of the scale time, increasing from top to bottom. For each combination of $v_{\rm kick}$ and $\tau$, we plot 100 snapshots of the density profile in units of the density at scale radius $\rho(r_s)$ at times spanning the simulation time, 0 to 200 scale times. Evolution at small radii is in part due to numerical artifacts; see text for further discussion.}
    \label{fig:density_profile_grid}
\end{figure}

In Fig.~\ref{fig:density_profile_grid}, we plot the halo density profiles with respect to the density $\rho(r_s)$ at the scale radius across the grid of kick velocities and lifetimes. For each panel, we overlay 100 snapshots spanning $0$ to $200\,t_s$ colored from purple (early) to yellow (late). In these figures, we include only particles that remain gravitationally bound to the halo. For small kick velocities and long lifetimes (bottom left), the evolution closely tracks that of a non-decaying halo, and the structural response to decay is small. Moving towards larger kick velocities and shorter lifetimes (up and to the right), the profile evolves more strongly in time. Once the kick velocity exceeds roughly a quarter of the scale velocity, the halo structure changes qualitatively. The profile becomes shallower at small radii and steeper at large radii; we quantify this decay-induced modification relative to the evolved control in Sec.~\ref{sec:emulator}, which removes the common numerical evolution of the inner profile.

\subsection{Velocity dispersions}

\begin{figure}[t]
    \centering
    \includegraphics[width=\linewidth]{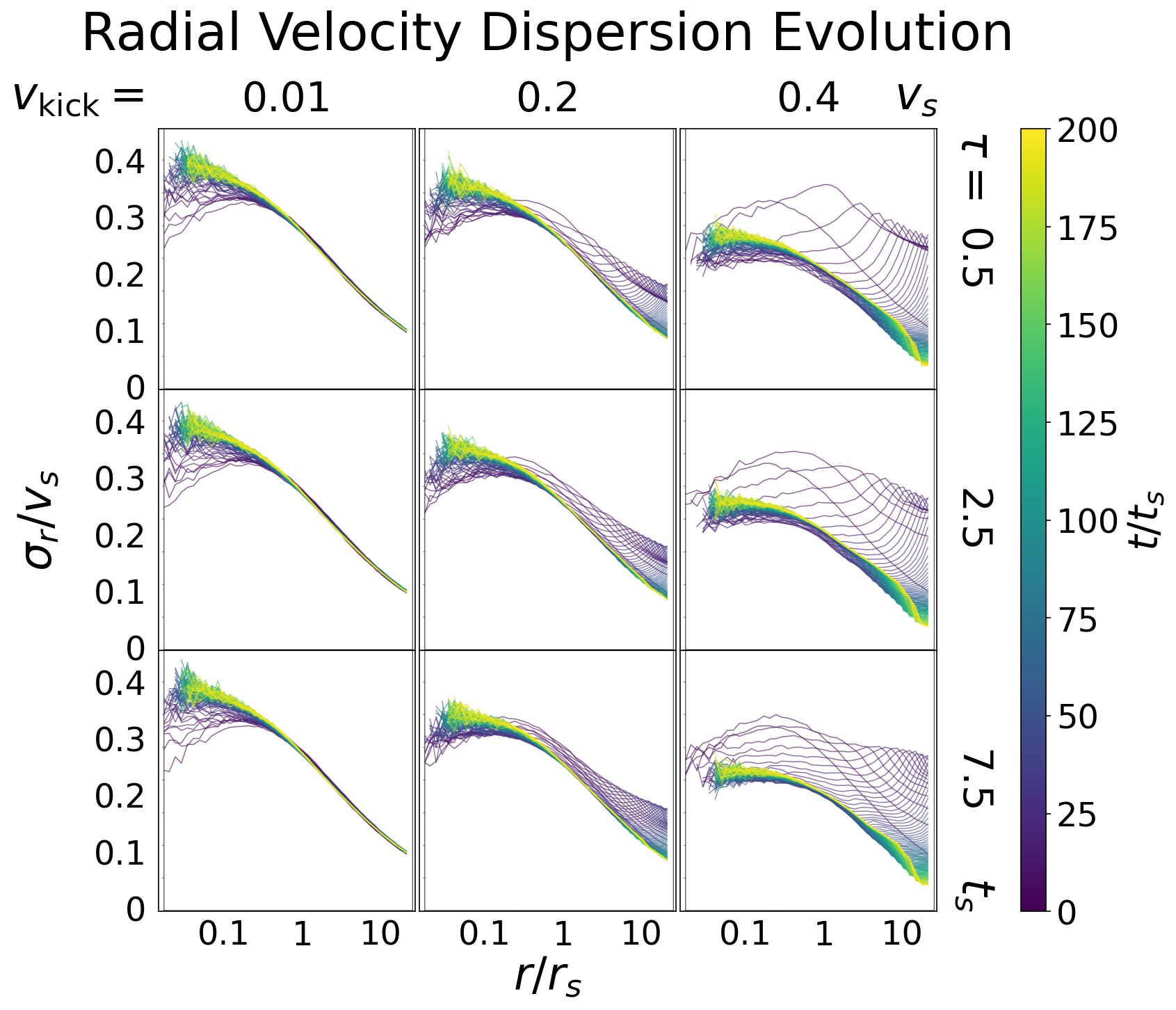}
    \caption{Radial velocity dispersion evolution of Hernquist profile halos with decaying dark matter. The kick velocity $v_{\rm kick}$ ranges from $0.01$ to $0.4$ in units of the scale velocity, increasing from left to right. The decay lifetime $\tau$ ranges from $0.5$ to $7.5$ in units of the scale time, increasing from top to bottom. For each combination of $v_{\rm kick}$ and $\tau$, we plot 100 snapshots of the velocity dispersion at times spanning the simulation time, 0 to 200 $t_s$.}
    \label{fig:velocity_dispersion_grid}
\end{figure}

Figure~\ref{fig:velocity_dispersion_grid} shows the radial velocity dispersion $\sigma_r$ over the same parameter grid and time range (again excluding particles that are no longer gravitationally bound). The velocity data mirror the density results. Nonstandard evolution becomes evident for $v_{\rm kick} \geq 0.2\,v_s$. At the largest kick velocities, the early-time (purple) curves reveal a population of decay products pushed to large radii, since the injected kinetic energy expands the halo. After $\sim 25t_s$, the dispersion settles into a new relaxed state (green and yellow curves) that, unlike the initial profile, is more cored (flattened) and has systematically lower velocities at small radii.

Across the grid, the halo continues to evolve for longer than the interval over which particles decay, particularly at large kick velocities. For $10^6$ particles and lifetimes of a few scale times, essentially all decays complete by $t \approx \tau \ln N = \tau \times 6\ln 10\approx 13.8\,\tau$, well inside our simulated window of 200 scale times, even for $\tau=10\, t_s$. We attribute this to the broad distribution of orbital energies produced by the kicks: a particle whose kick leaves it bound but with an energy close to zero is placed on a highly eccentric orbit with an apocenter far outside the original halo and an orbital time that may be far longer than $t_s$. The halo thus continues to phase-mix toward its new equilibrium long after the final decay.

\section{Anisotropic distributions}
\label{sec:anisotropy}

The initial conditions adopted heretofore assume an ergodic phase-space distribution function (DF), that is, one that depends on the phase-space coordinates only through the specific energy $\mathcal{E}$, and not on the angular momentum $L$. However, cosmological $N$-body simulations find that halos are mildly radially anisotropic, with anisotropy parameter $\beta=1-\sigma_t^2/\sigma_r^2$ (where $\sigma_t$ and $\sigma_r$ are the transverse and radial velocity dispersions, respectively) rising from $\approx 0$ near the center to $\approx 0.3$--$0.5$ near the virial radius \cite{Hansen:2004qs, Wojtak:2008mg}. Because a part of the DDM signature identified above is driven by particles on radial orbits, e.g., those carried to large apocenter or ejected entirely, the relaxed profile could in principle depend on the initial orbital anisotropy. In an effort to understand the sensitivity of the results to the ergodic assumption, we run simulations for models with several different values of $\beta$ \cite{Kamionkowski:2025uae}, restricting our attention here, for simplicity, to models with constant $\beta$. Here, $\beta = 0$ corresponds to an isotropic velocity distribution, $\beta > 0$ indicates radial anisotropy (orbits biased toward radial motion), and $\beta < 0$ indicates tangential anisotropy. For this comparison we fix the decay timescale and kick velocity to $\tau = 1\,t_s$ and $v_{\rm kick} = 0.4\,v_s$, and vary $\beta \in \{-0.4,\, 0.0,\, 0.4\}$.

\begin{figure}[h!]
    \centering
    \includegraphics[width=\linewidth]{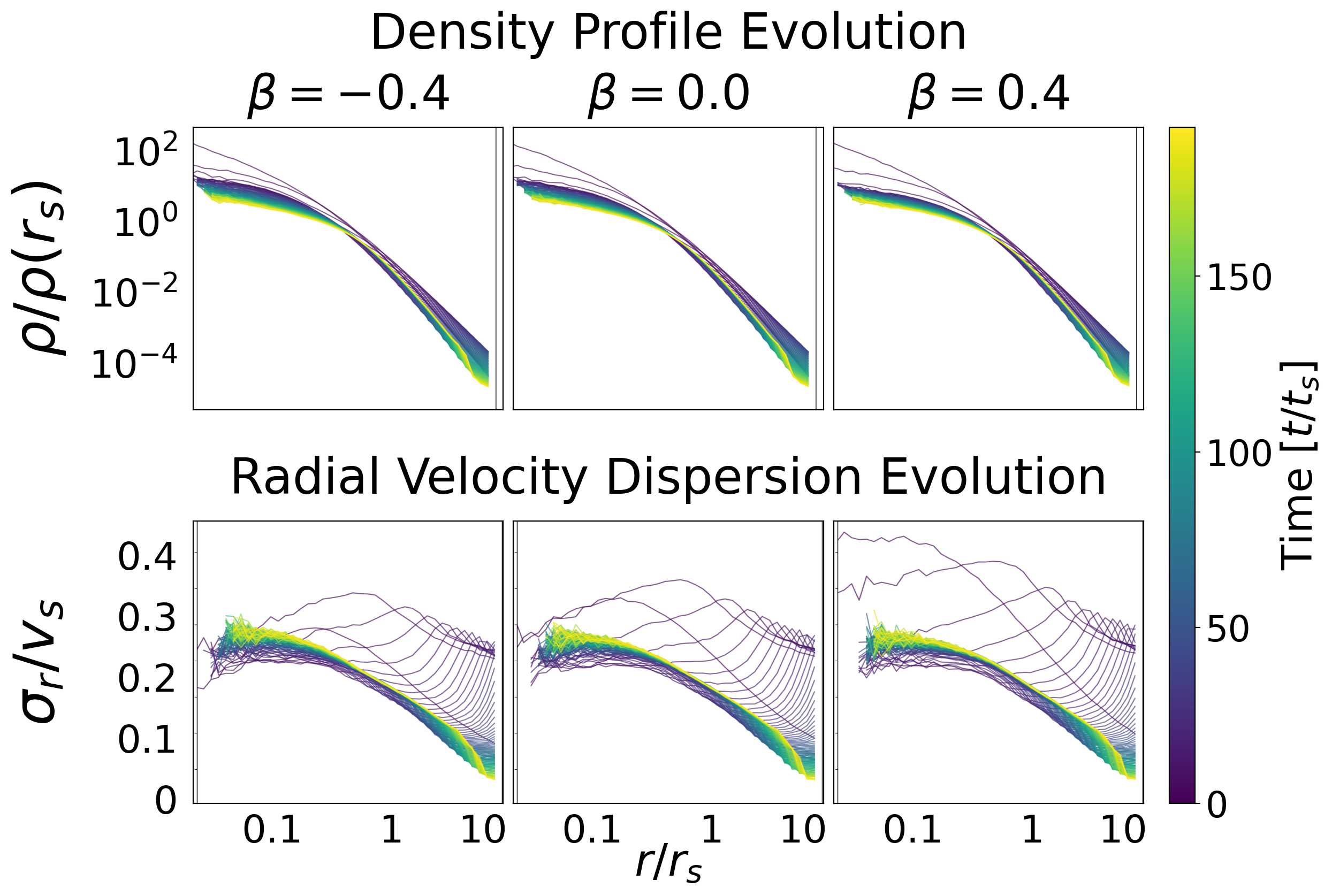}
    \caption{Evolution of a DDM halo with $\tau = 1.0\,t_s$, $v_{\rm kick} = 0.4\,v_s$, and three values of the anisotropy parameter, $\beta = -0.4, 0.0, 0.4$ (left to right). Each panel shows 100 snapshots of the relevant profile at times spanning the full simulation duration, 0 to 200 $t_s$. \textit{Top}--- density profile evolution. \textit{Bottom}--- radial velocity dispersion evolution.}
    \label{fig:beta_density}
\end{figure}

Figure~\ref{fig:beta_density} shows the time evolution of the density profile (top row) and the radial velocity dispersion (bottom row) for the three values of $\beta$. The density evolution is largely insensitive to the choice of anisotropy: the top row shows qualitatively similar profile evolution across all three columns. In contrast, the velocity dispersion evolution exhibits modest but systematic differences, discussed below.

For $\beta > 0$, the radial velocity dispersion is initially higher than for $\beta < 0$, reflecting the radial bias of the initial orbit population. At early times (purple curves), particles on eccentric radial orbits are preferentially ejected when combined with outward decay kicks. This produces the outward-moving density enhancement in the top row of Fig.~\ref{fig:beta_density}, corresponding to particles that decayed early and were subsequently expelled. However, a similar feature appears for all three values of $\beta$, indicating that this early-time ejection is driven primarily by the large kick velocity rather than the initial anisotropy. The thermalization of the velocity dispersion is consistent with previous studies of halo virialization in warm dark matter models \cite{Polisensky:2015eya}.

After this transient, the velocity dispersion profiles relax to similar configurations for all three $\beta$. At late times (yellow curves), the radial dispersion profiles are nearly indistinguishable, showing that decay dynamics erase the memory of the initial orbital anisotropy, consistent with the isotropic kick prescription. 

\begin{figure}
    \centering
    \includegraphics[width=\linewidth]{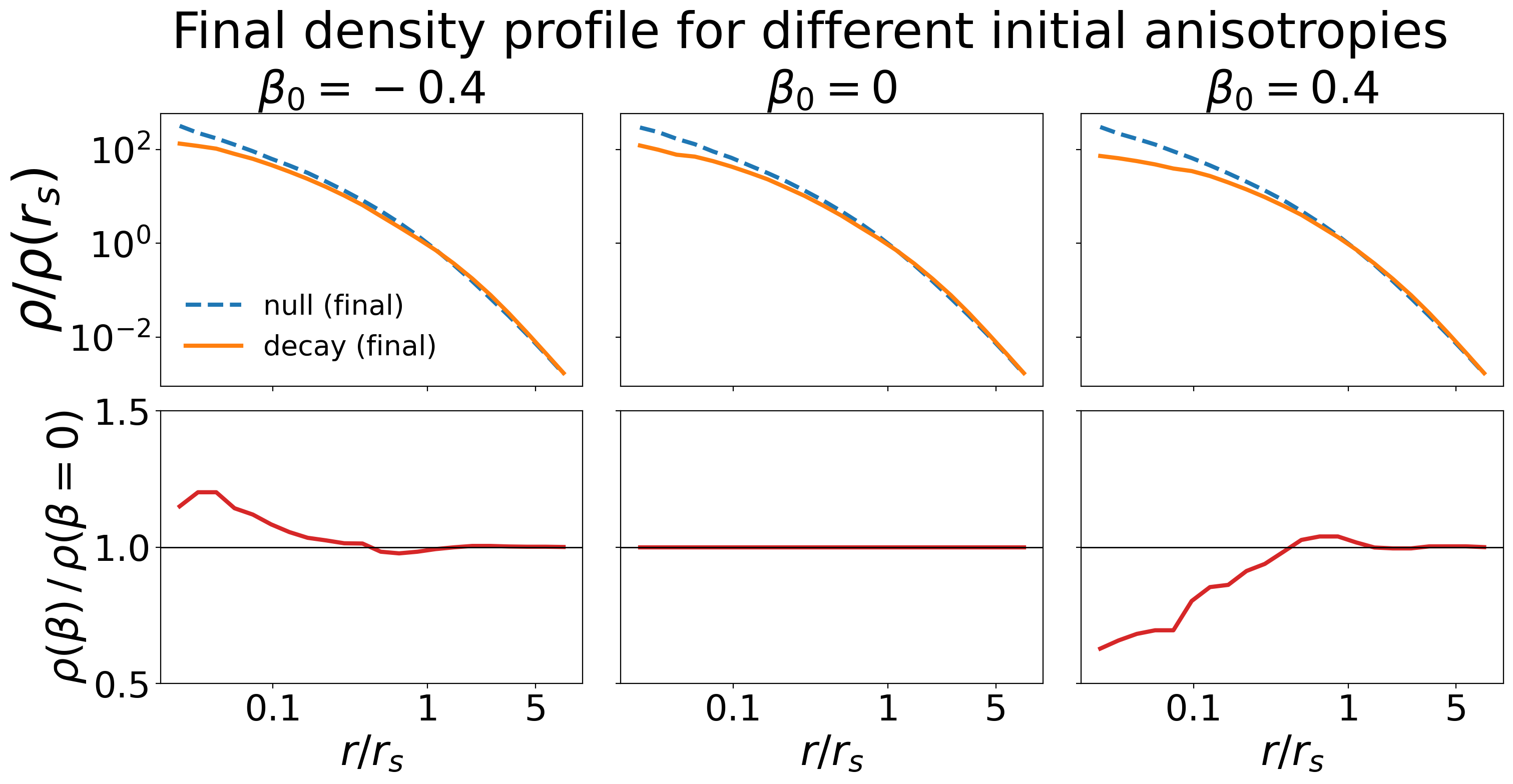}
    \caption{Final density profiles for DDM halos with different initial anisotropies (from left to right, $\beta_0 = -0.4,0.0,0.4$). \textit{Top}--- final density profiles for DDM halos (orange) and null halos (blue, dashed). \textit{Bottom}--- final density profile of DDM halos (red) relative to the $\beta_0 = 0$ case.}
    \label{fig:beta_final_density}
\end{figure}

One may also probe the sensitivity of the inner density profile to anisotropy. For a fixed density profile, the velocity dispersion required to hold the halo in equilibrium depends on the orbital anisotropy: radially biased systems ($\beta > 0$) must support themselves with a larger central dispersion than isotropic or tangentially biased ones ($\beta < 0$), since radial orbits contribute little pressure transverse to the radial direction. This is illustrated for the Hernquist model in Fig.~4.4 of \cite{binney2011galactic}, which shows the line-of-sight velocity dispersion profiles of a family of models that share the same density but differ in anisotropy, demonstrating that the central dispersion rises as the orbits become more radial. For a dark matter halo, we do not observe the projected dispersion directly. The more relevant quantity is the three-dimensional dispersion, which plays the role of an effective temperature for a collisionless halo. Still, the same ordering holds; a radially biased center is in this sense ``hotter'' than a tangentially biased one. One might therefore expect the tangentially biased (colder) centers to be more easily reshaped by kicks. For this analysis, we reduce the kick velocity to $v_{\rm kick} = 0.25v_s$ so that the effect of the kick does not totally evaporate the halo. We use $\tau = 1t_s$ as before and evolve the runs for 5 scale times. We match the decaying runs with non-decaying control halos at the same initial $\beta_0$.

Figure~\ref{fig:beta_final_density} shows the final density profiles. The inner density responds systematically and oppositely to the naive expectation. Relative to the matched null, the median density per radial bin within a scale radius is reduced to 74\% for $\beta_0 = -0.4$, 72\% for $\beta_0 = 0.0$, and 62\% for $\beta_0 = +0.4$. Relative to the $\beta = 0$ case, the density within a scale radius is increased to 105\% for $\beta_0 = -0.4$, and decreased to 86\% for $\beta_0 = +0.4$. With either comparison, the radially-biased halo is excavated the most, and the tangentially-biased halo the least.

We may understand this response by considering the angular momentum imparted by the decays. Each kick, and therefore angular momentum, is distributed isotropically in the rest frame, driving the velocity distribution towards isotropy. A radially-biased halo is dominated by low angular momentum orbits that plunge through the center, and is therefore most strongly affected. The added angular momentum has the effect of increasing the pericenters of these radial orbits so that they no longer reach small radii, thus carving out the inner density. In a tangentially-biased halo, the orbits already carry substantial angular momentum and large pericenters, making the inner structure comparatively resilient. The density signature is a counterpart of the velocity-space isotropization noted above and reinforces the conclusion that decay tends to erase the initial orbital anisotropy.

\begin{figure}
    \centering
    \includegraphics[width=\linewidth]{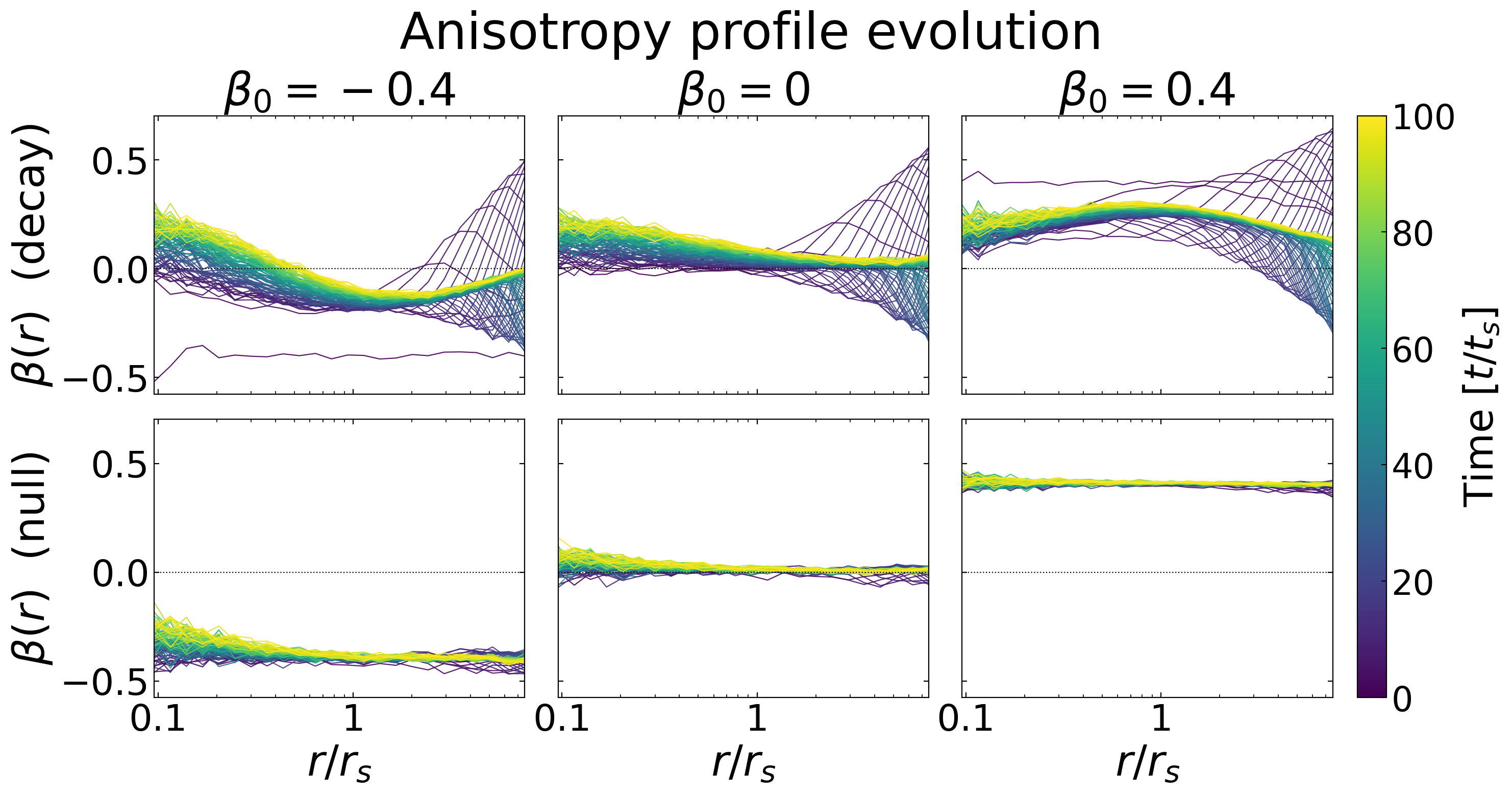}
    \caption{Evolution of the anisotropy parameter as a function of radius for halos evolved with $\beta_0 = -0.4$ (left), $\beta_0 = 0$ (middle), and $\beta_0 = +0.4$ (right). \textit{Top}---Halos with decays enabled, \textit{Bottom}--- null halos. Null halos on average retain their initial anisotropy, with some numerical relaxation at small radii for the case of $\beta_0 = -0.4$. Decay halos exhibit a transient period of highly radial distributions at large radii for times $t \lesssim 10\,t_s$, after which the anisotropy parameter settles towards a slightly radially biased state at small radii ($r < r_s$), while generally retaining its initial anisotropy at large $r$.}
    \label{fig:beta-decay-vs-null}
\end{figure}

Finally, one may consider the radial dependence of the anisotropy, and how it evolves for DDM vs CDM halos. Figure~\ref{fig:beta-decay-vs-null} overlays snapshots of $\beta(r)$ at several dynamical times (color bar) for decaying halos (top) and non-decaying halos (bottom), for the three different initial $\beta$ values. The null halos retain their initial $\beta(r)$ throughout, with a mild inward drift at small radii for $\beta_0 = -0.4$, which is consistent with the residual numerical relaxation discussed in Appendix~\ref{appendix:convergence}.

We see direct evidence that decay erases the memory of the initial orbital anisotropy in the $\beta(r,t)$ evolution in Fig.~\ref{fig:beta-decay-vs-null}. The decaying halos develop a pronounced radial dependence. At early times ($t \lesssim 10t_s$), the anisotropy in the outer radii is highly radially biased; the velocity kicks send a population of decay products outwards on radial, plunging orbits. This feature is transient and subsides after all particles have decayed, escaped, or phase-mixed. The halo then settles towards a final $\beta(r)$ that demonstrates a radial gradient in how much memory of $\beta_0$ is retained. The inner region ($r < r_s$) has a short orbital time and is reprocessed over many dynamical times, and thus relaxes to a common, mildly radially biased state independent of $\beta_0$. We find that, across the same grid of lifetimes and kick velocities, decaying halos with both isotropic and anisotropic initial conditions tend towards a common value $\beta \sim 0.15$, essentially independent of the decay lifetime (mirroring the lifetime independence of the enclosed-mass deficit we derive below, Sec.~\ref{subsec:emulator-mass}). The outer region has longer orbital times, and therefore retains the sign of its initial anisotropy, that is, the $\beta_0 = -0.4$ halo remains tangentially biased at $r \gtrsim r_s$, while the $\beta_0 = +0.4$ halo remains radially biased there. 

Although the anisotropy profile retains some memory of $\beta_0$ at large radii, the quantity our emulator predicts is the density profile, which we show in Fig.~\ref{fig:beta_final_density} is remarkably insensitive to the initial anisotropy. Further, cosmological halos exhibit only mild central radial anisotropy. We therefore build our emulator on isotropic ($\beta_0=0$) initial conditions and note that including other initial conditions would be a simple extension. 

\section{Density-profile emulator}
\label{sec:emulator}

The simulations described in the preceding sections provide the late-time density profile of a decaying dark matter halo at a discrete set of decay lifetimes and kick velocities. We construct an emulator to interpolate the relaxed density profile throughout the sampled two-dimensional parameter space, $\rho_{\rm DDM}(r,\,\tau,\,v_{\rm kick})$, where $\tau$ is expressed in units of the scale time and $v_{\rm kick}$ is expressed in units of the scale velocity. The emulator is used only for interpolation within the domain covered by the simulations.

\subsection{Simulation sample and numerical reference profile}
\label{subsec:emulator-data}

The training sample consists of 100 simulations arranged on a $10\times10$ grid in decay lifetime and kick velocity, with $0.5\,t_s \leq \tau \leq 10\,t_s$ and $0.1\,v_s \leq v \leq 0.5\, v_s$. We restrict the emulator to $v_{\rm kick}\geq 0.1\,v_s$: below this, the kick energy is negligible against the local binding energy, and thus the density modification is negligible. The emulator is likewise restricted to $r\gtrsim 0.1\,r_s$, the radius above which the simulated profiles are numerically converged (Appendix~\ref{appendix:convergence}). All simulations are initialized from the same Hernquist halo with ergodic DF and are evolved with $10^6$ particles and run to $t_{\rm f}=200\,t_s$. We use $10^5$ timesteps for decaying halos, and $10^6$ timesteps for the control halo (Appendix~\ref{appendix:convergence}). The final particle distributions are binned in logarithmically spaced radial shells.
\begin{figure}
    \centering
    \includegraphics[width=\linewidth]{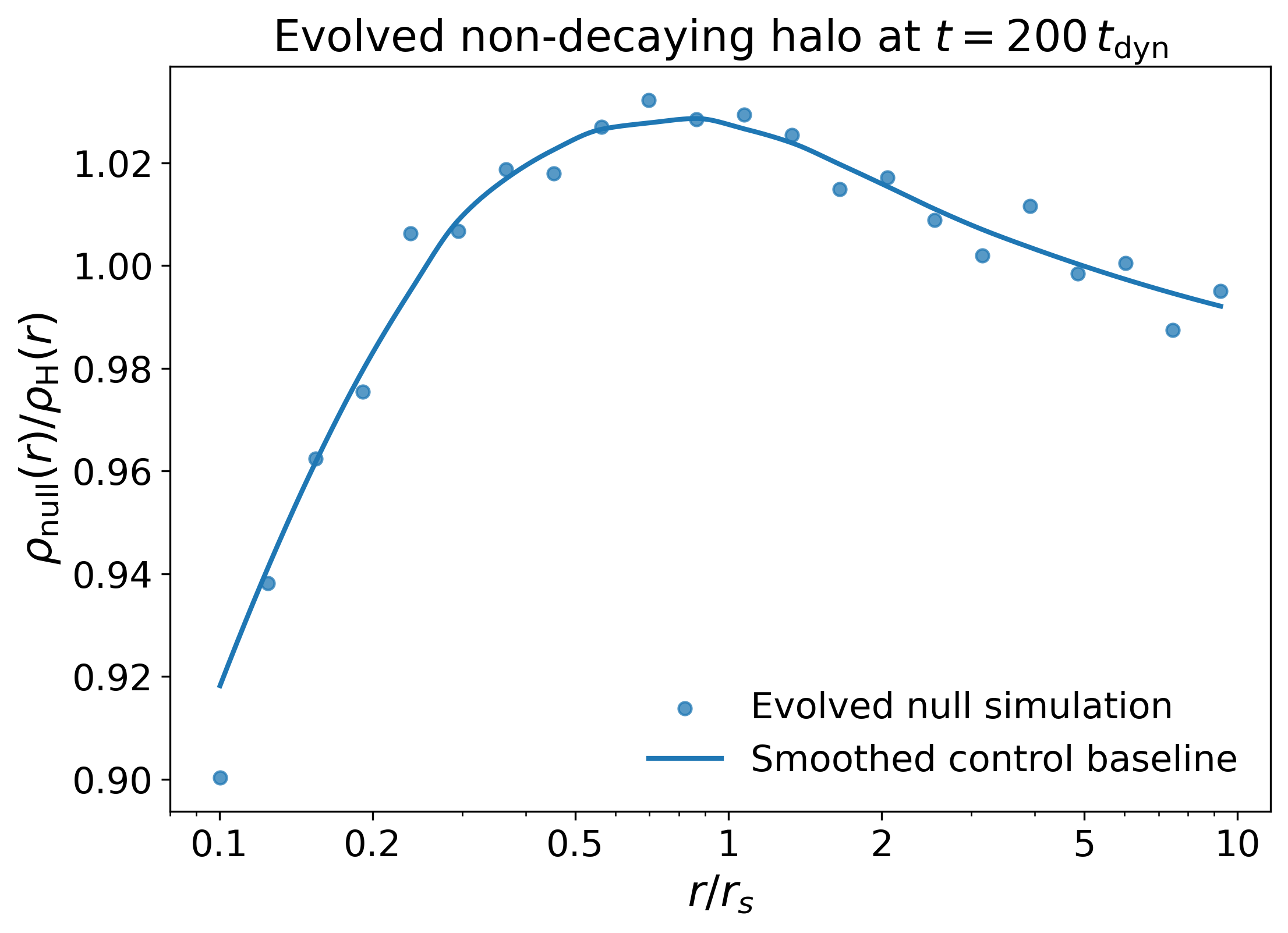}
    \caption{Density of the evolved non-decaying control halo at $t=200\,t_s$ relative to the analytic Hernquist profile used for the initial condition. Points show the binned control simulation, while the solid curve shows the smoothed baseline used to normalize the DDM density  profiles in the emulator.}
    \label{fig:null-baseline}
\end{figure}

As Fig.~\ref{fig:null-baseline} shows, numerical evolution alone redistributes density relative to the analytic initial condition $\rho_{\rm H}(r)$, most noticeably in the inner halo. A profile fit confirms that the evolved control halo remains Hernquist-like at large radii but develops a modestly shallower inner logarithmic slope than the initial $r^{-1}$ cusp; the central region is mildly cored relative to the analytic initial condition. Because this numerical evolution is common to the decaying and non-decaying runs, we normalize by the evolved control $\rho_{\rm null}(r)$ rather than by $\rho_{\rm H}(r)$, so that the emulated quantity isolates the physical effect of decay.

We retain all radial bins and weight each by its inverse Poisson variance, $\sigma^2(r) \approx 1/N_{\rm DDM}(r) + 1/N_{\rm null}(r)$ (with a small systematic floor). The principal-component decomposition and the mean profile are computed with these weights.

\subsection{Emulated quantity}
\label{subsec:emulated-quantity}

Rather than emulating the absolute density directly, we model the logarithmic density modification relative to the evolved control halo,
\begin{equation}
    y(r,\,\tau,\,v_{\rm kick})
    \equiv
    \ln\left[
        \frac{
            \rho_{\rm DDM}(r,\,\tau,\,v_{\rm kick})
        }{
            \rho_{\rm null}(r)
        }
    \right].
    \label{eq:emulator-target}
\end{equation}
This definition separates the additional structural response caused by dark matter decay from numerical evolution common to all simulations. It also ensures that the reconstructed density remains positive.

Each simulated profile is first interpolated onto a common logarithmic radial grid. We then use an inverse-variance-weighted principal-component analysis to express the radial dependence as
\[
    y(r,\,\tau,\,v_{\rm kick})
    =
    \overline{y}(r)
    +
    \sum_{i=1}^{N_{\rm PC}}
    c_i(\tau,v_{\rm kick})\,e_i(r),
\]
where $\overline{y}(r)$ is the mean log-density modification over the training sample, $e_i(r)$ are orthogonal radial principal components, and $c_i(\tau,v_{\rm kick})$ are their amplitudes. The final emulator retains $N_{\rm PC} = 4$ principal components, which captures the response to better than percent-level accuracy. Each amplitude is interpolated across the parameter plane by an independent Gaussian process with an anisotropic Mat\'ern-$5/2$ kernel in $(\ln\tau,\ln v_{\rm kick})$, plus a white-noise term for regularization. The absolute density follows as
\begin{equation}
    \rho_{\rm emu}(r,\,\tau,\,v_{\rm kick})
    =
    \rho_{\rm null}(r)
    \exp\left[
        \overline{y}(r)
        +
        \sum_{i=1}^{4}
        c_i(\tau,v_{\rm kick})\,e_i(r)
    \right].
    \label{eq:emulator-final}
\end{equation}
For use without the Gaussian-process machinery, each amplitude is also provided as a closed-form polynomial fit, $c_i(\tau,v_{\rm kick})=\sum_{j+k\le3}a^{(i)}_{jk}\,\tau^{\,j}\,v_{\rm kick}^{\,k}$, which through Eq.~\eqref{eq:emulator-final} reproduces the full emulator to a median fractional error of $0.7\%$ in density, which is well within its own $\sim1\%$ accuracy. The trained emulator, the radial functions $\overline{y}(r)$, $e_i(r)$, and $\rho_{\rm null}(r)$, the polynomial coefficients, and a ready-to-use evaluation routine are publicly available at \url{https://github.com/annabencke1/DDM_emulator}.

\subsection{Model selection and validation}
\label{subsec:emulator-validation}
\begin{figure}
    \centering
    \includegraphics[width=\linewidth]{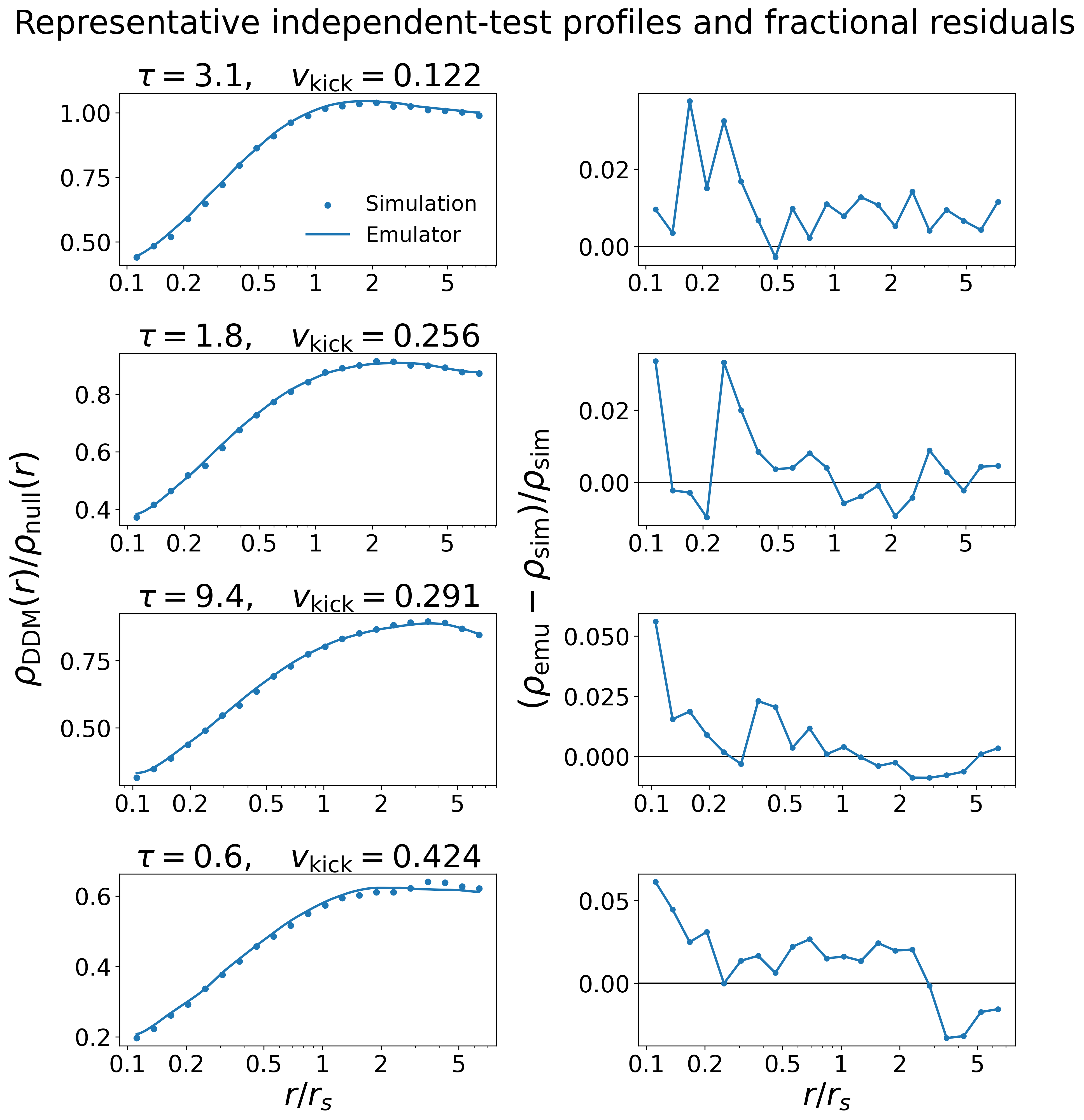}
    \caption{Representative predictions for simulations in the independent test set. The left panels show the simulated density modification relative to the evolved non-decaying control halo, $\rho_{\rm DDM}(r)/\rho_{\rm null}(r)$, together with the emulator prediction. The right panels show the corresponding dimensionless fractional residual, $(\rho_{\rm emu}-\rho_{\rm sim})/\rho_{\rm sim}$. The values of the decay lifetime $\tau$ and kick velocity $v_{\rm kick}$ are listed above each case in units of the scale time and scale velocity, respectively. The simulations shown here were not used during emulator construction or model selection.}
    \label{fig:emulator-tests}
\end{figure}
The emulator architecture is selected using only the training grid via boundary-aware cross-validation, withholding simulations from the interior of the grid and from the low-kick, short-lifetime, and long-lifetime regions. For each cross-validation fold, the PCA decomposition and Gaussian-process interpolation are refit without the withheld simulations, and the excluded density profiles are then predicted.

We quantify accuracy with the fractional density error, $(\rho_{\rm emu} - \rho_{\rm sim})/\rho_{\rm sim}$. Across the cross-validation sample, the emulator reproduces the withheld profiles with a median fractional error of $0.7\%$ and a median 90th-percentile error of $1.8\%$. Refitting on the full grid and evaluating on the 18 independent-test halos gives a median fractional error of $1\%$ and a 90th-percentile error of $2\%$ over the validated domain $0.1\,\lesssim r/r_s \lesssim 7.5$. This domain is bounded below by an independent timestep-convergence criterion (Appendix~\ref{appendix:convergence}). All radial bins within the domain are retained and weighted by their inverse Poisson variance.

Representative independent-test predictions are shown in Fig.~\ref{fig:emulator-tests}. The emulator reproduces both the overall radial dependence of the DDM modification and the absolute density profiles with percent-level typical accuracy. The largest residuals are generally confined to a small number of radial bins rather than appearing as coherent discrepancies across the full profile. We provide all independent-test predictions in Appendix~\ref{appendix:full-test-profiles}.

\subsection{Derived enclosed-mass profiles}
\label{subsec:emulator-mass}

For many applications, a compact description of the residual halo's bulk mass is more convenient than the full profile. We define a fixed aperture $R_{\rm ap} \equiv 8\,r_s$ and $M_{\rm ap} \equiv M(<R_{\rm ap})$. The aperture is close to the virial radius for our choice of halo mass and radius, and encloses $\simeq 80\%$ of the initial halo mass. Thus, $M_{\rm ap}$ serves as a virial-mass analogue without reference to $\rho_{\rm crit}$. We compute $M_{\rm ap}$ by directly integrating the simulated profiles, allowing the aperture to extend slightly beyond the quoted emulator range.

For the 100 relaxed profiles on the $(\tau,v_{\rm kick})$ grid, decay-induced heating reduces the enclosed mass according to
\begin{equation}
    \frac{M_{\rm ap}(\tau, v_{\rm kick})}{M_{{\rm ap},0}}
        = 1 - A\,v_{\rm kick}^{\,a}\,\tau^{b} \approx 1 - 2.2v_{\rm kick}^2,
    \label{eq:aperture-fit}
\end{equation}
where $v_{\rm kick}$ is in units of the scale velocity, $\tau$ in scale times, and $M_{{\rm ap},0}$ is the non-decaying control value. The best-fit parameters are $A = 2.2,\,a = 1.9,\,b = 0.02$.
Eq.~\eqref{eq:aperture-fit} reproduces the simulated $M_{\rm ap}$ with a median fractional accuracy of $0.3\%$ and a maximum deviation of $1.6\%$ across the sampled domain.

The weak lifetime dependence, $b \simeq 0.02$, arises partly because by $200,t_s$ essentially all particles have decayed, making the total injected kinetic energy nearly independent of $\tau$ and the mass deficit primarily controlled by $v_{\rm kick}$. The residual $\tau$ dependence reflects decay timing relative to halo relaxation, consistent with Sec.~\ref{sec:results}. Likewise, $a \simeq 1.9$ is close to the $a=2$ scaling expected from kinetic-energy injection $\propto v_{\rm kick}^2$; the small offset reflects kicked particles remaining bound at larger radii. Fixing $a=2$ still reproduces $M_{\rm ap}$ to $\simeq1.5\%$.

\section{Conclusion}
\label{sec:conclusions}

We have employed the \texttt{NSphere} code to simulate the dynamical evolution of spherically symmetric dark-matter halos composed of particles that decay to slightly less massive daughters with an associated velocity kick. Our simulations span a range of decay lifetimes and kick velocities, demonstrating that halos with short lifetimes and large kick velocities undergo significant structural changes, including core formation and expansion as in \cite{Nadler:2025yni}. We analyze the effect of anisotropic initial conditions, finding that the final profiles of DDM halos tend to become radially biased regardless of initial anisotropy. The effect of decays is most pronounced on radially biased initial conditions due to the isotropization of angular momentum.

We constructed an emulator based on principal-component analysis and Gaussian-process regression that accurately interpolates the relaxed density profiles across our parameter space, achieving median fractional errors of $\sim1\%$ over the validated radial domain. Because it maps $(\tau, v_{\rm kick})$ directly onto a full radial density profile at negligible cost, the emulator is well suited for insertion into halo-model and halo occupation distribution frameworks, where the per-halo DDM response can be propagated to statistical observables such as galaxy-galaxy lensing profiles, the subhalo mass function, and satellite abundances \cite{Nadler:2024ims,Peter:2010sz,Abdelqader:2008wa, Ando:2021fhj, Gilman:2026uvq, Montandon:2026jyl, DES:2022doi}. This offers a route to connect the nonlinear, halo-scale signatures computed here to the large-scale observables that DESI, Planck, and weak-lensing surveys already constrain \cite{FrancoAbellan:2021, Bucko:2024izb, Bencke:2026uws}.

We identify a structural response in halo density profiles that becomes appreciable at $v_{\rm kick} \gtrsim 0.1\,v_s$, which for a $10^{12}\,M_\odot$ halo is $\approx 40\,{\rm km\,s^{-1}}$, adjacent to the bounds derived by linear probes. The free-streaming of decay products is constrained by the suppression of the matter power spectrum and from the Lyman-$\alpha$ forest, with constraints ranging from $v_{\rm kick} \lesssim 10^{-2}c$ (CMB and BAO) \cite{FrancoAbellan:2021, Montandon:2025xpd, Bencke:2026uws} to $v_{\rm kick} \lesssim 100\,\rm{km\,s}^{-1}$ (Lyman-$\alpha$ forest) \cite{Wang:2013rha, Wang2014, Fuss:2022zyt, Zhao:2026wxi, DES:2022doi, Peter:2010sz, refId0}. The next generation of surveys will either tighten these bounds or probe an independent corner of the model. In particular, the Rubin Observatory, Euclid, and the Roman Space Telescope will resolve the inner density slopes of thousands of dwarf galaxies and detect the substructure-induced perturbations of cold stellar streams \cite{Mao:2022fyx}, both of which are sensitive to the cored profiles and modified velocity dispersions predicted here.

Future work will extend this analysis to NFW and Einasto profiles, incorporate the halo-mass dependence through concentration-mass relations \cite{Peter_2010, Peter2010b}, and fold the emulated profiles into forward models of dwarf-galaxy rotation curves and stellar-stream gaps to derive quantitative, halo-scale constraints on the DDM parameter space. 

\appendix
\section{Full test profiles}
\label{appendix:full-test-profiles}

\begin{figure*}[t]
    \centering
    \includegraphics[width=\linewidth]{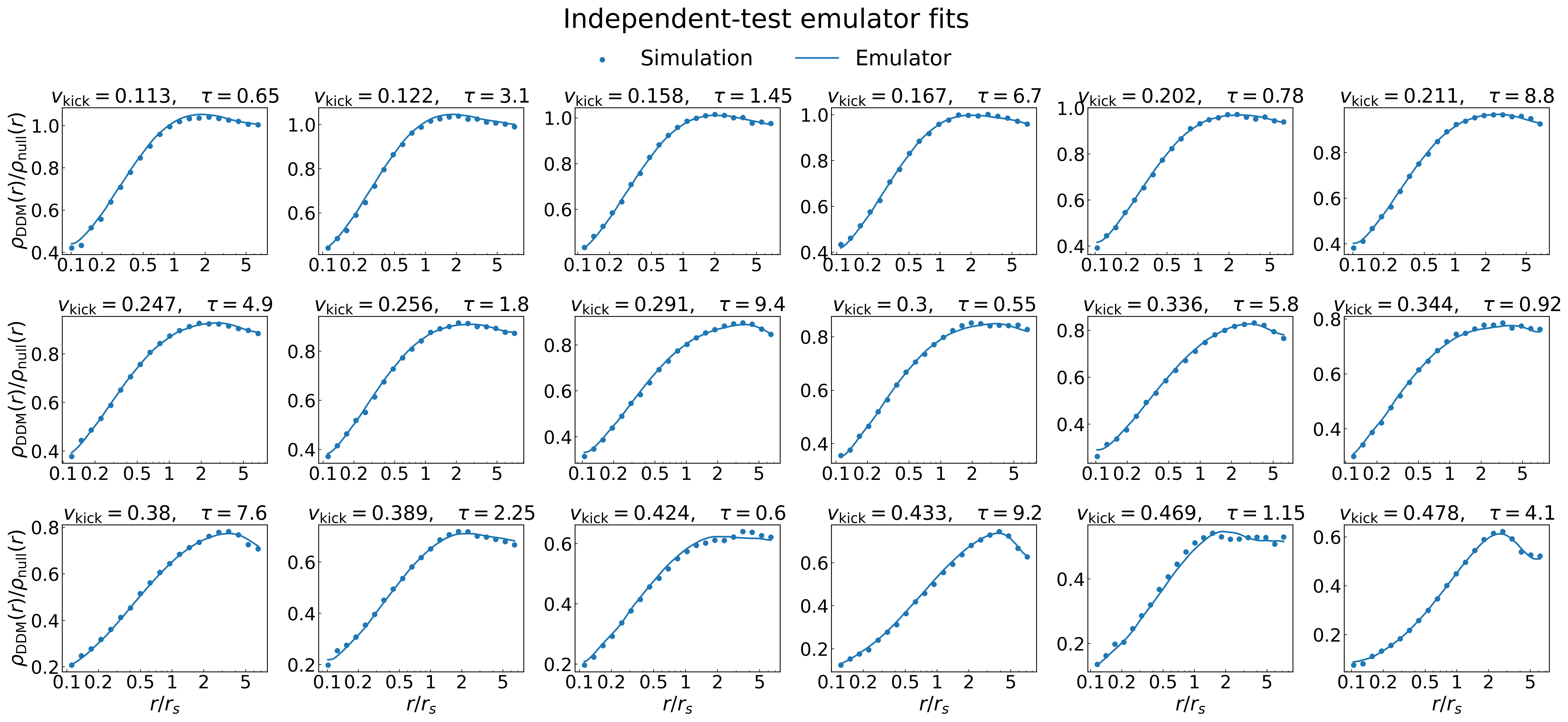}
    \caption{Complete set of independent test profiles used to validate the emulator. In each panel, points show the simulated density modification relative to the evolved non-decaying halo, $\rho_{\rm DDM}(r)/\rho_{\rm null}(r)$, and the solid curve shows the corresponding emulator prediction. Each panel is labeled by its decay lifetime $\tau$ (in scale times) and its kick velocity $v_{\rm kick}$ (in units of the scale velocity). The emulator reproduces the simulated profiles at percent-level accuracy across the sampled parameter space, with the largest residuals confined to isolated radial bins rather than coherent offsets.}
    \label{fig:all-test-profiles}
\end{figure*}

Figure~\ref{fig:all-test-profiles} presents the complete set of eighteen independent test profiles used to validate the emulator. Each panel shows the simulated density modification relative to the evolved control halo alongside the emulator prediction.

\section{Numerical convergence}
\label{appendix:convergence}

\begin{figure}
    \centering
    \includegraphics[width=\linewidth]{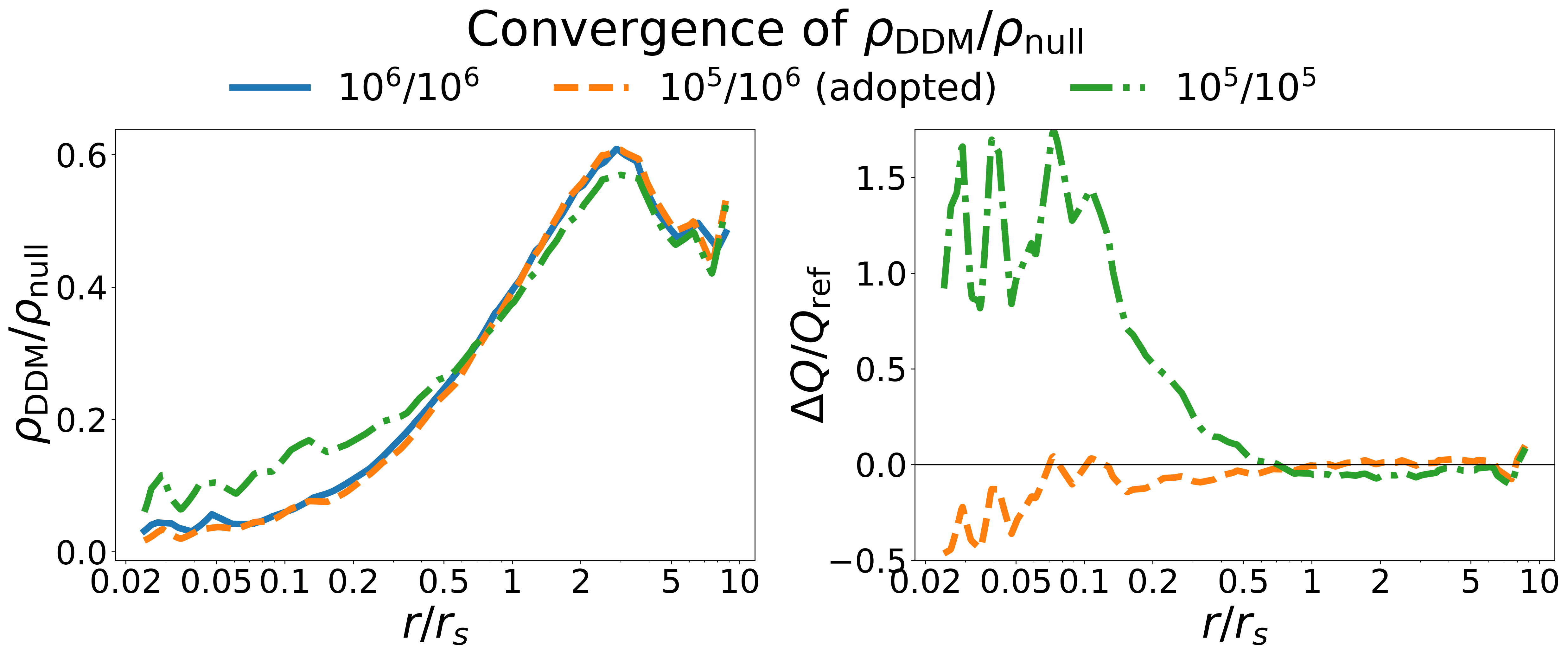}
    \caption{Numerical convergence of the normalized density profile $\rho_{\rm DDM}/\rho_{\rm null}$ for a representative halo ($\tau=5\,t_s$, $v_{\rm kick}=0.25\,v_s$). \textit{Left}---the ratio at three time resolutions: the fully converged reference (both runs at $10^{6}$ timesteps, solid), the adopted mixed resolution (decaying run at $10^{5}$, control at $10^{6}$, dashed), and the naive equal-resolution case (both at $10^{5}$, dot-dashed). \textit{Right}--- fractional difference of $Q \equiv \rho_{\rm DDM}/\rho_{\rm null}$ each from the $10^{6}/10^{6}$ reference. The adopted $10^{5}/10^{6}$ normalization tracks the converged reference to better than $\sim10\%$ over the entire validated domain $r\gtrsim0.1\,r_s$, whereas the equal-resolution $10^{5}/10^{5}$ ratio is biased by up to $\sim140\%$ in the inner halo, where the unconverged control cusp inflates the ratio. The bias is systematic rather than statistical, and motivates both the mixed-resolution normalization and the inner radial cut at $0.1\,r_s$.}
    \label{fig:convergence}
\end{figure}

The central density of a Hernquist halo evolved with a fixed-step integrator will naturally experience numerical relaxation. This affects the decaying and non-decaying runs unequally. In the decaying halo, the energy injected by the kicks flattens the cusp into a physical core, which the integrator holds accurately; we find that the decaying profile is already converged at $10^5$ timesteps. The non-decaying control halo retains its steep cusp, which is far more sensitive to time resolution and remains under-resolved at $10^5$ steps. 

This asymmetry dictates our normalization. We evolve the decaying grid at $10^5$ timesteps and normalize by a control run evolved at $10^6$ timesteps, thus dividing a converged decaying profile by a converged reference. Fig.~\ref{fig:convergence} shows that this adopted mixed-resolution ratio $\rho_{\rm DDM}(10^5)/\rho_{\rm null}(10^6)$ tracks the fully converged $\rho_{\rm DDM}(10^6)/\rho_{\rm null}(10^6)$ to better than $\sim 10\%$ (median a few percent) across the validated domain. The naive equal-resolution ratio $\rho_{\rm DDM}(10^5)/\rho_{\rm null}(10^5)$ is instead biased by up to $\sim 140\%$ in the inner halo, spuriously inflating the ratio and flattening the inferred core. Equal but insufficient time resolution is therefore not a substitute for convergence.

Below $\simeq 0.1\,r_s$, the adopted ratio still retains a residual bias from the innermost decaying bins, and this sets the inner radius of the emulator-validated domain. We emphasize that this is a systematic bias, not shot noise; it cannot be addressed by the inverse-variance weighting applied within it.

\begin{acknowledgments}
MK was supported by NSF Grant No. 2412361, NASA ATP Grant No.\ 80NSSC24K1226, and the Templeton Foundation. AP was supported by NASA Grant No.\ 80NSSC26K0221. 
\end{acknowledgments}
 

\small
\bibliography{ddmdynamics}{}

\end{document}